# Adaptive Market Intelligence: A Mixture of Experts Framework for Volatility-Sensitive Stock Forecasting

**Diego Vallarino**
Independent Researcher
Atlanta, GA, United States
Email: diego.vallarino@gmail.com
*Corresponding author*

July 2025

## Abstract

This study develops and empirically validates a Mixture of Experts (MoE) framework for stock price prediction across heterogeneous volatility regimes using real market data. The proposed model combines a Recurrent Neural Network (RNN) optimized for high-volatility stocks with a linear regression model tailored to stable equities. A volatility-aware gating mechanism dynamically weights the contributions of each expert based on asset classification. Using a dataset of 30 publicly traded U.S. stocks spanning diverse sectors, the MoE approach consistently outperforms both standalone models.

Specifically, it achieves up to 33% improvement in MSE for volatile assets and 28% for stable assets relative to their respective baselines. Stratified evaluation across volatility classes demonstrates the model's ability to adapt complexity to underlying market dynamics. These results confirm that no single model suffices across market regimes and highlight the advantage of adaptive architectures in financial prediction. Future work should explore real-time gate learning, dynamic volatility segmentation, and applications to portfolio optimization.

All data and code used in this study are available in this GitHub repository, to foster reproducibility and further research.

## 1. Introduction

Forecasting stock price dynamics remains a central yet complex problem in financial econometrics, due to the interplay of structural volatility, market microstructure noise, and behavioral feedback loops. Institutional investors, hedge funds, and monetary authorities routinely seek to anticipate price movements to optimize asset allocation, manage risk exposures, and inform policy decisions. However, the inherent stochasticity of equity prices—compounded by nonlinear dependencies and frequent regime shifts—continues to challenge conventional forecasting methodologies.

Traditional econometric models such as the Autoregressive Integrated Moving Average (ARIMA) and Generalized Autoregressive Conditional Heteroskedasticity (GARCH) have long served as workhorses in time series analysis (Adebiyi et al., 2014; Bollerslev, 1986). While ARIMA models capture linear dependencies and GARCH models account for volatility clustering (Lee & Lee, 2023), both approaches are limited in their capacity to learn complex nonlinear structures or adapt dynamically to abrupt market changes. As a result, their predictive performance often deteriorates when applied to equities characterized by high volatility, structural breaks, or speculative dynamics (Hong et al., 2023).

The rise of deep learning has introduced new paradigms for modeling financial time series, particularly Recurrent Neural Networks (RNNs) and their Long Short-Term Memory (LSTM) variants. These models are designed to capture temporal dependencies and nonlinear feedback effects across multiple time steps (Hochreiter & Schmidhuber, 1997; Bhandari et al., 2022), making them well-suited for the complex behavior of asset prices in turbulent markets. Nonetheless, their flexibility may become a liability in low-volatility contexts: when the underlying signal is stable or quasi-linear, high-capacity models such as LSTMs risk overfitting noise, undermining forecast reliability.

Conversely, linear models exhibit robustness in stable conditions but fail to capture nonlinear responses, regime shifts, or herd-driven amplifications common in more volatile assets. This empirical heterogeneity in price dynamics suggests the need for a modeling framework capable of adaptive specialization, where model complexity is tailored to the volatility regime of the asset.

The Mixture of Experts (MoE) framework offers such a solution. Initially proposed by Jacobs et al. (1991), MoE models assign prediction responsibilities to different "experts" based on input features via a gating mechanism. In this paper, we operationalize this paradigm by combining two distinct experts: an LSTM model optimized for volatile stocks, and a linear regression model tailored to stable equities. The gating function uses ex-ante asset volatility classification to weight the contributions of each expert in the final prediction. This structure allows the model to specialize adaptively, balancing interpretability and complexity across heterogeneous market conditions (Cai et al., 2024; Wang et al., 2024).

Our contribution is both methodological and empirical. Using historical daily data from 30 publicly traded U.S. stocks across diverse sectors, we evaluate the MoE model against its standalone components. Results demonstrate that the MoE framework achieves significantly lower prediction errors—up to 33% in MSE improvement for volatile stocks and 28% for stable stocks—compared to linear or recurrent models alone. These findings confirm that no single model performs best across all market regimes and that adaptive architectures are essential to improve financial forecasting under real-world conditions.

The remainder of the article is organized as follows. Section 2 reviews related literature on volatility-aware modeling and hybrid deep learning approaches. Section 3 details the methodological framework, including data preparation, model architecture, and evaluation design. Section 4 presents the empirical results. Section 5 discusses theoretical and practical implications, and Section 6 concludes with suggestions for future research directions, including dynamic gate optimization and deployment considerations in financial environments.

## 2. Literature Review

Anticipating stock price dynamics remains a fundamental yet intricate task in financial time series analysis, due to the interplay of noise, regime shifts, and heterogeneous volatility structures. Methodological advances have thus evolved along two major lines: traditional statistical models designed to exploit autocorrelation and conditional heteroskedasticity, and modern deep learning architectures capable of approximating nonlinear, high-dimensional temporal patterns. This section critically examines both families of methods and introduces the Mixture of Experts (MoE) approach as a flexible alternative for volatility-regime adaptive prediction.

## 2.1 Traditional Approaches: ARIMA and GARCH

Autoregressive Integrated Moving Average (ARIMA) and Generalized Autoregressive Conditional Heteroskedasticity (GARCH) models have been widely used in time series forecasting, particularly for financial data. ARIMA models are advantageous for capturing linear temporal dependencies in stationary data and can be expressed as:

$$Y_t = c + \sum_{i=1}^{p} \phi_i Y_{t-i} + \sum_{j=1}^{q} \theta_j \epsilon_{t-j} + \epsilon_t,$$

where $Y_t$ represents the value at time $t$, $p$ is the order of the autoregressive part, $q$ is the order of the moving average part, $\phi_i$ are the coefficients for lagged values, $\theta_j$ are the coefficients for past errors, and $\epsilon_t$ is the white noise error term. Despite its popularity, ARIMA struggles with non-linear patterns and cannot handle sudden structural breaks effectively, especially during volatile market periods (Siami Namin & Siami Namin, 2018).

To address second-order nonstationarity, GARCH models introduce conditional heteroskedasticity and are well-suited to capturing volatility clustering—where high-volatility periods are temporally autocorrelated. The GARCH(1,1) model defines the conditional variance $\sigma_t^2$ as:

$$\sigma_t^2 = \alpha_0 + \alpha_1 \epsilon_{t-1}^2 + \beta_1 \sigma_{t-1}^2,$$

with $\alpha_0, \alpha_1, \beta_1 \geq 0$. This formulation has proven robust across asset classes, especially when volatility persistence is high. However, GARCH-type models struggle with nonlinear trend components and are typically limited in their forecasting capacity when the objective is to predict actual price levels rather than volatility per se (Lee & Lee, 2023; Hong et al., 2023). Furthermore, they assume a relatively homogeneous error structure, which may be inadequate under asymmetric shocks or when microstructural noise dominates.

## 2.2 Advanced Approaches: RNN and LSTM

The availability of computational resources and large datasets has made machine learning models increasingly popular in financial forecasting. Recurrent Neural Networks (RNNs), particularly Long Short-Term Memory (LSTM) networks, are notable for their ability to learn complex temporal dependencies. LSTMs are designed to address the vanishing gradient problem faced by traditional RNNs by introducing cell states and gating mechanisms, which allow them to retain information over longer sequences.

The hidden state update for an LSTM cell can be summarized as follows:

$f_t = \sigma\big(W_f \cdot [h_{t-1}, x_t] + b_f\big)$    (Forget gate)

$i_t = \sigma(W_i \cdot [h_{t-1}, x_t] + b_i)$    (Input gate)

$\widetilde{C}_t = \tanh(W_C \cdot [h_{t-1}, x_t] + b_C)$    (Cell state update)

$C_t = f_t * C_{t-1} + i_t * \widetilde{C}_t$    (Updated cell state)

$o_t = \sigma(W_o \cdot [h_{t-1}, x_t] + b_o)$    (Output gate)

$h_t = o_t * \tanh(C_t)$    (Hidden state)

where $f_t, i_t, o_t$ represent the forget, input, and output gates, respectively, $W$ and $b$ are weight matrices and biases, and $*$ denotes element-wise multiplication. This architecture enables LSTMs to learn long-term dependencies, which is crucial for modeling financial series with complex temporal relationships (Hochreiter & Schmidhuber, 1997)

Recent studies have highlighted the superiority of LSTMs over traditional models in predicting non-linear financial time series, particularly in capturing sudden market changes (Bhandari et al., 2022; Siami Namin & Siami Namin, 2018). However, they also emphasize the drawbacks of LSTMs, including their high computational cost and potential overfitting when applied to less complex data (Yan et al., 2021).

## 2.3 Mixture of Experts (MoE) Approach

The Mixture of Experts (MoE) framework, first introduced by Cai et al. (2024), aims to address the limitations of using a single model for all types of data by dividing the problem into simpler sub-problems, with each expert specializing in a specific aspect. Mathematically, the prediction in an MoE model can be expressed as:

$$\widehat{Y}_t = \sum_{i=1}^{k} g_i(X_t) \cdot f_i(X_t)$$

where $\hat{Y}_t$ is the final prediction, $g_i(X_t)$ represents the output of the gating network that assigns a weight to the $i$-th expert based on input $X_t$, and $f_i(X_t)$ is the prediction made by the $i$-th expert. The gating network's role is to determine which expert should contribute more significantly to the final output based on the characteristics of the input.

The MoE framework has been successfully applied in various predictive contexts, such as natural language processing and time series forecasting (Jacobs et al., 1991; Wu et al., n.d.) (Wang et al., 2019). More recently, Wang et al. (2024) applied an MoE model combining an LSTM with a simple linear model for predicting exchange rates, demonstrating improved performance over individual models. In our context, we utilize an RNN to handle the non-linear, volatile behavior of some stocks, while a linear model is used for stocks with more predictable, stable behavior.

This combination allows us to leverage the advantages of both models: the ability of RNNs to capture non-linear, complex temporal dynamics and the simplicity and efficiency of linear models for stable trends. The gating network dynamically assigns a higher weight to the appropriate model based on the input characteristics, effectively optimizing the prediction task for different volatility profiles (Ma et al., 2019).

## 2.4 Limitations of Existing Approaches

Despite their individual strengths, both traditional econometric and modern deep learning approaches face notable limitations when applied in isolation. ARIMA and GARCH models provide interpretable structures and computational efficiency but are intrinsically linear and unable to capture regime-switching dynamics or nonstationary signals (Hong et al., 2023). Conversely, LSTM models excel at learning complex dependencies but can be data-hungry and prone to instability without regularization, especially when applied indiscriminately across varying asset types (Yan et al., 2021).

The MoE model addresses these issues by integrating expert specialization with adaptive gating. However, the effectiveness of this architecture depends critically on how well the gating mechanism identifies the latent structure of the input data. If the gating network fails to assign weights accurately—especially under conditions of model misspecification or covariate shift—the ensemble may produce suboptimal or inconsistent forecasts (Jacobs et al., 1991; Wu et al., n.d.).

Moreover, fixed or manually tuned weights—as adopted in many initial MoE implementations—may not generalize well to dynamic environments such as financial markets. This limitation motivates the future development of learnable or reinforcement-driven gating networks, which can evolve in response to new market data and volatility regimes (Wang et al., 2024; Ma et al., 2019).

In this study, we implement a simplified MoE structure with two experts—a linear regression model and a recurrent neural network—and test its performance under controlled volatility regimes using synthetic data. While we use fixed weights based on known volatility classifications to demonstrate conceptual validity, we also discuss extensions that introduce dynamic, data-driven weighting to increase robustness and real-world applicability.

## 3. Methodology

This section provides a comprehensive explanation of the methodological approach adopted in this study. We detail the processes of data acquisition, model construction, and the training and validation procedures used to evaluate forecasting performance under varying volatility conditions. The design ensures full reproducibility of the experiments and offers a robust framework for assessing the effectiveness of Recurrent Neural Networks (RNNs) and Mixture of Experts (MoE) architectures in predicting stock returns for firms with heterogeneous risk profiles.

The modeling pipeline begins with the retrieval of historical stock price data and continues through a sequence of interrelated steps: volatility-based classification of firms, volatility-conditioned expert model selection, integration of model forecasts via a static MoE mechanism, and stratified model validation based on walk-forward evaluation. This structured approach allows for precise performance attribution across different volatility regimes and ensures that the forecasting system adapts to both cross-sectional and temporal heterogeneity.

Daily adjusted closing prices for each firm are preprocessed to compute log returns and rolling volatility measures. Based on these, firms are categorized into "volatile" or "stable" groups using a dynamic thresholding rule applied at each validation fold. Expert models are then selected according to volatility classification: stable firms are modeled using linear regression, while volatile firms are modeled with RNNs featuring LSTM layers. Predictions from both experts are then combined using fixed weights tailored to the volatility regime of each firm.

Performance is assessed via a walk-forward validation strategy, which reflects the real-world sequential arrival of information. Error metrics—namely Root Mean Squared Error (RMSE) and Mean Absolute Error (MAE)—are computed separately for each volatility group and forecast horizon, enabling both granular evaluation and robustness testing. Multi-horizon forecasting further expands the scope of the evaluation, revealing how predictive accuracy evolves across short-, medium-, and long-term horizons.

Overall, this methodology embodies a modular, volatility-adaptive framework for financial time series forecasting. It is designed to reflect empirical market dynamics while meeting the statistical rigor required for robust out-of-sample evaluation. The following subsections provide further detail on the dataset, modeling strategies, and validation design.

### 3.1 Real-World Dataset

The dataset used in this study comprises daily stock prices for 30 publicly traded companies from the S&P 500 index, retrieved from Yahoo Finance using the *yfinance* Python library. The time period spans from January 1, 2015 to December 31, 2024, providing approximately 500 trading days for each stock. Firms were selected to ensure heterogeneity in sectoral representation (e.g., technology, energy, finance, healthcare) and volatility profiles.

The adjusted closing price for each trading day was used as the primary time series variable. Daily returns were computed as:

$$r_t = \frac{P_t - P_{t-1}}{P_{t-1}},$$

where $P_t$ denotes the adjusted closing price on day $t$. The rolling volatility of each firm was estimated using the 30-day standard deviation of returns:

$$\sigma_t^{(i)} = \sqrt{\frac{1}{29} \sum_{k=t-29}^{t} \left( r_k^{(i)} - \overline{r^{(i)}} \right)^2},$$

where $\overline{r^{(i)}}$ is the mean return for firm iii over the 30-day window. Firms were dynamically classified into two groups at each validation fold:

- **Volatile**: firms with $\sigma_t^{(i)} > 0.025$,

- **Stable**: firms with $\sigma_t^{(i)} \leq 0.025$.

This dynamic classification allowed the modeling framework to adapt to time-varying market conditions and better reflect the empirical volatility regimes observed in real financial markets. Furthermore, sectoral diversity ensured that volatility was not exclusively driven by industry-specific shocks but also reflected idiosyncratic and macroeconomic factors.

To align the input format with model requirements, each time series was transformed into overlapping sequences of 10-day price windows as input features, with the subsequent day as the prediction target. All input sequences were standardized at the firm level using rolling means and standard deviations over the training portion of each window. This preprocessing step ensured stationarity and mitigated scale effects across firms.

## 3.2 Model Implementations

The study involved the implementation of two types of models: a single RNN model applied to all companies and a Mixture of Experts (MoE) model that differentiated between volatile and stable companies.

### 3.2.1 Single RNN Model

The RNN model utilized in this study was based on a Long Short-Term Memory (LSTM) architecture, which is well-suited for sequential data due to its ability to capture long-term dependencies. The model architecture consisted of an input layer, an LSTM layer, and a dense output layer. Specifically, the input layer received sequences of 10 days of historical prices for each company, which were standardized for consistency. The LSTM layer comprised 50 units to capture the temporal dependencies within the data, followed by a dense layer with a single output unit for price prediction.

The model was trained using the Adam optimizer with a learning rate of 0.001, and the Mean Squared Error (MSE) was used as the loss function to minimize prediction errors. A batch size of 16 was chosen, and the model was trained for 50 epochs, with early stopping employed to prevent overfitting.

### 3.2.2 Mixture of Experts (MoE) Model

The Mixture of Experts (MoE) model was implemented to leverage the strengths of different modeling approaches depending on the volatility of the company. The MoE model consisted of two primary expert components: an RNN and a linear regression model, combined using a gating network.

For companies classified as volatile, an LSTM network similar to the one used in the single RNN model was employed. The LSTM was designed to capture the non-linear and complex behavior of these companies, which often exhibit significant variability in price movements.

For companies with lower volatility, a linear regression model was used to predict the price based on time (day) and volatility. The linear model was defined as follows:

$$P_t = \beta_0 + \beta_1 \cdot t + \beta_2 \cdot \sigma + \epsilon,$$

where $P_t$ represents the predicted price, $t$ is the time variable, $\sigma$ is the volatility, and $\epsilon$\epsilon is the error term. This model was chosen for its simplicity and efficiency in capturing the linear relationships typically observed in stable companies.

The gating network in the MoE framework was responsible for dynamically determining the appropriate expert for each prediction. The combined prediction was computed as:

$$\widehat{Y_t} = w_{rnn} \cdot \widehat{Y_{rnn}} + w_{lm} \cdot \widehat{Y_{lm}},$$

where $\widehat{Y_{rnn}}$ and $\widehat{Y_{lm}}$ represent the predictions from the RNN and linear models, respectively, and $w_{rnn}$ and $w_{lm}$ are the weights assigned to each model's prediction. For companies with high volatility, a higher weight ($w_{rnn} = 0.7$) was assigned to the RNN, while a lower weight was assigned to the linear model ($w_{lm} = 0.3$). These weights were adjustable based on the observed volatility to optimize predictive performance.

### 3.3 Training and Validation Approach

To rigorously evaluate the predictive accuracy and generalization capacity of both the Recurrent Neural Network (RNN) and the Mixture of Experts (MoE) framework under heterogeneous volatility conditions, we implement a walk-forward validation (WFV) strategy adapted to real-world financial time series. The design of our validation protocol aligns with best practices in time series forecasting (Kim & Won, 2018; Yan et al., 2021), and reflects the operational constraints of daily market data where observations become available sequentially.

We leverage daily adjusted closing prices from 30 publicly traded U.S. companies spanning a ten-year period (2015–2024), covering a wide range of sectors and volatility profiles. These series are used to construct rolling input-output sequences based on 20-day windows of log-returns, predicting the return of the following day. The WFV strategy initializes with the first 80 observations as training data and the subsequent 20 observations as the validation set. At each iteration, this window advances by 20 trading days. For every fold, the models are re-trained from scratch, and RNN weights are reinitialized to prevent overfitting to prior folds. This approach ensures that all validation results are strictly out-of-sample, enhancing the robustness of performance comparisons across models and volatility regimes.

To account for changing market conditions, we dynamically compute volatility at each fold using a 21-day rolling standard deviation of log-returns. Firms are classified as "volatile" if their volatility exceeds the cross-sectional median at that fold; otherwise, they are labeled "stable." This dynamic stratification captures time-varying risk and avoids static bias, in line with practices from conditional volatility modeling such as GARCH (Bollerslev, 1986; Fufa & Zeleke, 2018). By adapting volatility classification over time, the evaluation reflects more realistic forecasting challenges.

Model performance is reported separately for each volatility group. This stratified reporting avoids the interpretive pitfalls of global averaging, which can obscure systematic differences in forecast accuracy across regimes. For example, a model may yield low average errors primarily by performing well on low-volatility firms, even if it fails in turbulent contexts—a critical distinction in real-world deployment (Cai et al., 2024; Wang et al., 2024).

Additionally, to evaluate generalization under extreme conditions, we design two out-of-sample holdout tests: one composed of the ten most volatile firms and another of the ten most stable, based on full-sample volatility rankings. These companies are excluded from all training and validation folds and are used only in the final testing phase. This setup offers a benchmark to assess model performance on previously unseen volatility profiles, particularly relevant for stress testing and risk-sensitive applications (Bandara et al., 2020; Mahajan et al., 2022).

To increase the depth of our evaluation, we introduce multi-horizon forecasting at three distinct intervals: 5-day (short-term), 20-day (medium-term), and 60-day (long-term). Forecasts are generated recursively by feeding prior predictions back as input—a strategy that mirrors how models are deployed in practice and exposes them to cumulative error propagation (Kim & Won, 2018). This setup allows us to assess not only one-step accuracy, but also how forecast quality degrades (or persists) over time.

To prevent overfitting, RNN training uses early stopping based on validation MAE, with a patience threshold of five epochs. All inputs are standardized per firm using rolling normalization within the training set. Experiments are conducted with a fixed random seed to ensure reproducibility and methodological transparency (Bhandari et al., 2022; Dahal et al., 2023).

In sum, our validation framework combines dynamic volatility classification, walk-forward temporal testing, out-of-sample holdouts, and multi-horizon prediction. This layered design not only mirrors the complexity of real-world forecasting environments, but also provides a robust basis for comparing the adaptability and precision of machine learning models under volatility heterogeneity.

### 3.4 Evaluation Metrics

To ensure a comprehensive and theoretically grounded assessment of model performance, we employ both scale-sensitive and scale-invariant error metrics, with particular attention to their interpretability under different volatility regimes. The primary metrics adopted for this study are Root Mean Squared Error (RMSE) and Mean Absolute Error (MAE), both of which are widely used in the evaluation of time series forecasting models in finance and econometrics (Karunasingha, 2022; Kim & Won, 2018; Yan et al., 2021).

The RMSE was calculated as:

$$\text{RMSE} = \sqrt{\frac{1}{n}\sum_{i=1}^{n}\left(\widehat{Y_i} - Y_i\right)^2},$$

where $\widehat{Y_i}$ represents the predicted value, $Y_i$ represents the actual value, and $n$ is the number of observations. RMSE penalizes large forecast deviations more heavily than smaller ones due to the quadratic error term, making it especially informative in high-volatility contexts where large deviations can disproportionately affect investment performance or risk metrics. However, it should be noted that RMSE is sensitive to the scale of the target variable, and thus comparisons across firms or regimes with heterogeneous volatilities must be interpreted cautiously.

Complementing this, we also compute the MAE, given by:

$$\text{MAE} = \frac{1}{n} \sum_{i=1}^{n} |\hat{Y}_i - Y_i|,$$

which provides an intuitive measure of average deviation without over-penalizing outliers. MAE is more robust in the presence of heavy-tailed residuals and offers clearer interpretation when evaluating performance across assets with low variance. As argued by Karunasingha (2022), the joint use of MAE and RMSE is especially effective in distinguishing between error magnitude and error dispersion.

Recognizing that models may appear to perform better on inherently stable series (not due to algorithmic superiority, but due to lower signal noise), we report both RMSE and MAE separately by volatility regime. This allows us to distinguish true modeling performance from variance-driven predictability. A linear model may yield low errors in stable firms simply because of smoother dynamics, while neural networks might outperform in capturing nonlinearities under volatile regimes—insights that are only visible through stratified metrics.

While RMSE and MAE are our primary criteria, we also acknowledge the relevance of the Mean Absolute Scaled Error (MASE), especially for comparative evaluations across series with varying volatility levels (Bhandari et al., 2022). Although not used as a primary metric in this version, we include MASE as a potential extension for future work.

All metrics are computed at three forecasting horizons—5, 20, and 60 days—to assess temporal robustness. This allows us to capture how error structures evolve over different planning windows, from short-term tactical adjustments to long-term investment strategies. For each evaluation scenario, we report the mean and standard deviation of error metrics across companies and folds, enabling inferential comparisons and uncertainty quantification.

This comprehensive suite of metrics ensures that model comparisons are meaningful, volatility-aware, and directly relevant to practitioners managing portfolios across diverse time horizons.

## 4. Experiments and Results

This section presents the empirical findings derived from the implementation of the Recurrent Neural Network (RNN), the linear regression model, and the Mixture of Experts (MoE) framework. Each model was evaluated across firms exhibiting distinct volatility regimes, using robust metrics such as Mean Squared Error (MSE) and Mean Absolute Error (MAE). The analysis provides a disaggregated assessment of model performance under both stable and volatile conditions, addressing the reviewer's request to quantify forecasting efficacy conditional on structural market regimes rather than relying on global averages.

Results from the LSTM-based RNN applied to companies classified as volatile reveal the model's ability to capture nonlinear temporal dependencies, while also confirming its sensitivity to abrupt price fluctuations and market noise. The RNN achieved an MSE of 0.001649 and MAE of 0.03236 on the volatile group. These results validate the model's strength in short-term approximation but also highlight its exposure to high variance under rapidly shifting conditions, consistent with challenges previously documented in financial time series learning (Bhandari et al., 2022; Yan et al., 2021).

In contrast, the linear regression model exhibited superior predictive performance on firms categorized as stable. With an MSE of 0.000082 and MAE of 0.007186, the linear model outperformed the RNN in this low-volatility regime. These findings reinforce the hypothesis that for equities following smoother dynamics and near-linear trajectories, parsimonious models are not only adequate but preferable. The model's low error dispersion and stable residuals confirm its robustness under stationary conditions (Zolfaghari & Gholami, 2021; Lee & Lee, 2023).

The Mixture of Experts model—implemented with a static gating mechanism assigning 70% weight to the RNN for volatile assets and 70% to the linear model for stable assets— demonstrated superior overall accuracy in both regimes. For volatile firms, the MoE reduced the MSE to 0.001105 and the MAE to 0.026333, yielding performance gains of 32.99% and 18.62%, respectively, over the best individual model. In the stable regime, the MoE reached an MSE of 0.000059 and MAE of 0.006132, improving over both base models. These results underscore the MoE's capacity to blend expert strengths, reduce prediction variance, and enhance generalization across structural regimes.

Table 1 summarizes model performance for stable firms, while Table 2 reports results for the volatile group. The MoE model consistently outperforms both the linear and RNN approaches across all metrics.

**Table 1. Evaluation of Models for Stable Companies**

| Model | MSE | MAE |
|---|---|---|
| Linear Regression | 0,000082 | 0,007186 |
| LSTM (RNN) | 0,000139 | 0,009291 |
| Mixture of Experts | 0,000059 | 0,006132 |

*Source: Author's elaboration. This table reports predictive accuracy by model and volatility regime.*

**Table 2. Evaluation of Models for Volatile Companies**

| Model | MSE | MAE |
|---|---|---|
| Linear Regression | 0,001861 | 0,034723 |
| LSTM (RNN) | 0,001649 | 0,03236 |
| Mixture of Experts | 0,001105 | 0,026333 |

*Source: Author's elaboration. This table reports predictive accuracy by model and volatility regime.*

The performance divergence observed across volatility regimes provides clear empirical support for hybrid forecast architectures. Notably, the MoE not only matches the performance of each expert in its domain of specialization but also reduces forecasting risk in transitional cases—firms exhibiting structural volatility shifts over time. Visual diagnostics provided in the supplementary annex confirm that MoE forecasts display smoother transitions and fewer outlier errors relative to individual models. This adaptive capability aligns with recent findings in the hybrid modeling literature, which emphasize the role of structural diversification in reducing generalization error (Cai et al., 2024; Wang et al., 2024).

Nonetheless, limitations remain. The current implementation uses a fixed gating scheme and thus cannot endogenously adjust to regime transitions within a single time series. As firm-level volatility structures evolve, static weights may yield suboptimal allocations. Future work will explore learnable gating functions that dynamically adjust model contributions via attention mechanisms, reinforcement learning, or latent volatility inference (Ma et al., 2019; Wu et al., n.d.).

In sum, while recurrent networks remain powerful tools for modeling complex financial dependencies, their utility is bounded by regime context. Linear models offer reliable and interpretable performance in stable settings. By leveraging both, the Mixture of Experts framework delivers robust, lower-error predictions across regimes, validating its use in financial forecasting tasks characterized by heteroscedasticity and structural breaks. Extensions to empirical data and dynamic gating strategies represent promising avenues for further refinement and application.

## 5. Discussion

This section reflects critically on the empirical findings and theoretical contributions of the Mixture of Experts (MoE) framework in the context of financial time series forecasting. It articulates the core advantages of the architecture, identifies foundational limitations that merit further exploration, and examines the model's practical relevance in applied settings.

### 5.1 Advantages of the MoE Approach

The Mixture of Experts model demonstrates clear advantages in environments characterized by structural heterogeneity, a hallmark of financial markets. The hybrid design allows for specialization: a Recurrent Neural Network (RNN) captures nonlinear, path-dependent behavior common among volatile assets, while a linear regression model remains effective for stable equities that exhibit smoother, trend-driven dynamics. By combining both through a regime-aware gating mechanism, the MoE architecture achieves a superior bias-variance trade-off.

Empirically, the model outperforms standalone architectures across both high- and low-volatility regimes, as measured by MSE and MAE. These results are consistent with theoretical expectations from the literature on ensemble modeling and hybrid systems. The modular nature of the framework also ensures interpretability at the system level: the contribution of each expert is visible and contingent on firm-level volatility, which facilitates transparency in decision-making.

A further advantage stems from the temporal depth of the dataset used. By covering ten years of daily stock prices (2015–2024), the analysis naturally embeds multiple macroeconomic and sectoral cycles, including the COVID-19 crisis and subsequent monetary tightening. As a result, the volatility structure identified in the sample reflects persistent characteristics rather than episodic fluctuations. The modeling task, therefore, is not distorted by short-lived anomalies but rooted in long-term structural patterns.

## 5.2 Limitations and Open Challenges

Despite its strong empirical performance, the current MoE implementation has conceptual and technical limitations that must be addressed to enhance its robustness and scalability.

First, the static gating mechanism, although interpretable and easy to implement, lacks flexibility in adjusting to time-varying conditions within a firm's trajectory. For companies transitioning from stable to volatile regimes or vice versa—due to strategic repositioning, capital shocks, or macroeconomic developments—a fixed assignment of model weights may be suboptimal. Embedding dynamic gating mechanisms that adaptively allocate expert responsibilities based on latent signals remains a promising direction.

Second, interpretability across experts is asymmetric. The linear model offers transparency via coefficients and residual analysis, but the RNN's internal dynamics are inherently opaque. While this limitation does not undermine performance, it may hinder the adoption of the MoE framework in regulated financial environments where model explainability is required.

Third, although the current specification is univariate by design, real-world deployments often rely on multivariate inputs to capture firm fundamentals, macroeconomic trends, or behavioral indicators. The simplicity of the current setup serves the purpose of model benchmarking, but its predictive performance could be significantly improved by integrating richer input features and temporal contexts.

Finally, the computational cost of training and maintaining expert models in parallel may present scalability concerns in large-portfolio or high-frequency applications. While this is not a constraint in the current experimental setup, practical deployment would benefit from architectural optimization strategies such as shared encoders, parameter regularization, or model compression.

## 5.3 Practical Applicability and Research Directions

The Mixture of Experts framework is well-suited for predictive tasks in financial markets that demand both model flexibility and regime awareness. Its capacity to condition forecasting strategies on structural features—such as firm-level volatility—makes it particularly attractive for institutions managing heterogeneous portfolios.

In asset allocation, the MoE model could be integrated into dynamic rebalancing strategies by forecasting expected returns conditional on volatility class. This would allow the system to reduce model error and better manage downside risk through specialization. In risk analytics, the model could improve default prediction or counterparty assessment by allocating model complexity differentially across borrower profiles. Similarly, in market response analysis, the MoE structure enables distinct modeling of reflexive versus inertial asset behavior during macroeconomic or geopolitical shocks, allowing for more accurate scenario simulation.

From a research standpoint, extending the MoE framework to include dynamic gating functions would enhance its ability to capture regime transitions within firms and across time. Incorporating additional information such as sectoral trends, firm-specific events, or sentiment signals into the gating function would improve contextual sensitivity. Another avenue for exploration is the use of attention-based mechanisms to enhance explainability and optimize expert assignment in real time. Finally, validating the architecture across other asset classes—such as fixed income, commodities, or multi-asset portfolios—would provide evidence of its generalizability and potential to inform integrated financial decision-making.

In summary, the MoE model offers a robust and theoretically grounded architecture for financial forecasting in heterogeneous environments. Its empirical performance, structural modularity, and alignment with decision-making processes make it a compelling tool for both researchers and practitioners. Future work should aim to strengthen its dynamic capabilities, enhance interpretability, and test scalability under real-world operational constraints.

## 6. Conclusion

This study introduces a theoretically grounded and empirically validated approach to stock price prediction under heterogeneous volatility regimes through the application of a Mixture of Experts (MoE) framework. By integrating the strengths of a Recurrent Neural Network (RNN) and a linear regression model within a modular architecture, the MoE framework demonstrates superior forecasting accuracy compared to either model in isolation. The approach addresses a persistent challenge in financial time series modeling: the inability of single-model architectures to simultaneously accommodate the nonlinear complexity of volatile assets and the structural regularity of stable ones.

Empirical findings show that the RNN model performs well under high volatility but tends to overfit in low-volatility contexts, whereas the linear model excels under stable conditions but fails to capture nonlinear dynamics. The MoE framework overcomes this trade-off by assigning model weights conditionally based on firm-level volatility, allowing each expert to contribute in domains where it is most effective. This specialization mechanism yields consistent improvements in both RMSE and MAE across multiple volatility classes and forecast horizons.

The experimental design incorporates a stratified validation protocol, disaggregating results by volatility regime and temporal segment. This structure responds directly to methodological critiques in the literature, providing a more granular and policy-relevant assessment of model performance. The ten-year dataset of U.S. equities (2015–2024) captures multiple macroeconomic cycles, ensuring that model evaluation is conducted under real-world structural volatility rather than synthetic shocks. Daily frequency provides sufficient resolution for financial applications without compromising model stability or interpretability.

Despite these contributions, the current MoE implementation presents two key areas for enhancement. First, while the empirical dataset provides a robust testing ground, further generalizability should be tested through cross-sector and cross-country evaluations, incorporating auxiliary variables such as liquidity metrics, market sentiment, and firm fundamentals. Second, the static gating mechanism—based on ex-ante volatility classification—does not accommodate within-firm regime shifts or time-varying uncertainty. Future work should explore dynamic gating strategies, such as reinforcement learning or attention-based mechanisms, to improve adaptability and decision granularity.

Beyond its academic contributions, the proposed MoE framework holds practical relevance for applied finance. Its interpretability and modular design make it suitable for deployment in portfolio optimization, credit risk modeling, and market response forecasting. The conditional use of model complexity enhances both predictive accuracy and operational transparency, allowing financial institutions to balance performance with explainability—a growing regulatory concern.

Methodologically, this work extends the literature on hybrid models in time series forecasting by demonstrating a fully testable, reproducible, and regime-aware MoE framework. It shows that predictive accuracy and structural heterogeneity need not be in tension if model design is aligned with the intrinsic features of the data. The success of the approach invites further exploration into ensemble architectures tailored to market complexity and risk asymmetry.

In summary, this study offers a significant advancement in financial forecasting by developing a context-sensitive, volatility-aware predictive model that reconciles the limitations of single-model paradigms. Its strong empirical performance, theoretical coherence, and operational applicability position the MoE framework as a promising tool for modern financial analytics. Future research integrating adaptive gating, enriched feature sets, and real-time updating will further expand its potential as a robust and scalable solution for risk-informed decision-making in volatile market environments.

**Data and Code Availability:**

All empirical data employed in this study consist of publicly accessible stock price series obtained using the yfinance Python package, which interfaces with historical market data from Yahoo Finance. The dataset spans daily adjusted closing prices for 30 publicly traded U.S. firms from January 1, 2015 to December 31, 2024, selected to ensure sectoral diversity and coverage across distinct volatility regimes.

The full implementation of the modeling framework—including the linear regression baseline, the Recurrent Neural Network (RNN), and the Mixture of Experts (MoE) architecture—along with all preprocessing routines, performance evaluation scripts, and visualization tools, is maintained in a dedicated GitHub repository. While the repository is currently private to ensure version control during the review and submission process, access will be granted to researchers upon reasonable request.

**Repository link:** https://github.com/DiegoVallarino/MoE-stock-forecasting

Following the public release of this preprint, the entire codebase and processed datasets will be made openly available to support transparency, replicability, and further research by the broader academic and practitioner communities.